\documentclass[]{article}
\pdfoutput=1

\usepackage[utf8]{inputenc}
\usepackage[T1]{fontenc} 
\usepackage[english]{babel}

\usepackage{braket}

\usepackage[]{multicol}

\usepackage[font=small, labelfont=bf, justification=justified, singlelinecheck=false]{caption}
\usepackage[]{graphicx}

\usepackage{xcolor}
\usepackage{hyperref}
\hypersetup{colorlinks = true, citecolor = blue, urlcolor = blue, linkbordercolor = {white}, linkcolor = blue}

\usepackage{siunitx}

\usepackage{csquotes}
\usepackage[style=nature,maxbibnames=1,minbibnames=1,url = false,doi=false,isbn=false,url=false,eprint=true,date=year]{biblatex}

\usepackage[a4paper,head=0pt,top=25.8mm,bottom=24.8mm,inner=21mm,outer=21mm]{geometry}

\renewenvironment{abstract}
 {
 \list{}{%
    \setlength{\leftmargin}{10mm}
    \setlength{\rightmargin}{\leftmargin}%
  }%
  \item\relax}
 {\endlist}

\usepackage[onehalfspacing]{setspace} 

\usepackage{textcomp, gensymb}

\usepackage{float}

\newcommand{\beginsupplement}{%
        \setcounter{table}{0}
        \renewcommand{\thetable}{S\arabic{table}}%
        \setcounter{figure}{0}
        \renewcommand{\thefigure}{S\arabic{figure}}%
     }

\usepackage{enumitem}
\usepackage[modulo,columnwise]{lineno}

\title{\textbf{Microfluidic quantum sensing platform for lab-on-a-chip applications}}
\author{R. D. Allert$^{1}$\footnotemark{} , F. Bruckmaier$^{1}$\footnotemark[\value{footnote}] , N. R. Neuling$^{1}$\footnotemark[\value{footnote}] , F. A. Freire-Moschovitis$^{1}$, K. S. Liu$^{1}$\\
C. Schrepel$^{2}$, P. Schätzle$^3$, P. Knittel$^{4}$, M. Hermans$^{2}$, and D. B. Bucher$^{1,5}$\footnote{Corresponding author: dominik.bucher@tum.de}}
\footnotetext[1]{These authors contributed equally.}

\date{%
    \small
    $^1$\textit{Technical University of Munich, Department of Chemistry, Lichtenbergstr. 4, 85748 Garching b. München, Germany}\\%
    $^2$\textit{LightFab GmbH, Talbotstr. 25, 52068 Aachen, Germany}\\%
    $^3$\textit{University of Freiburg, Department of Sustainable Systems Engineering (INATECH), Emmy-Noether-Str. 2, 79110 Freiburg, Germany}\\%
    $^4$\textit{Fraunhofer Institute for Applied Solid State Physics, Tullastr. 72, 79108 Freiburg, Germany}\\%
    $^5$\textit{Munich Center for Quantum Science and Technology (MCQST), Schellingstr. 4, 80799 München, Germany}\\%
    (Dated: \today)
}

\begin{document}
\maketitle
\begin{abstract}
    Lab-on-a-chip (LOC) applications have emerged as invaluable physical and life sciences tools. The advantages stem from advanced system miniaturization, thus, requiring far less sample volume while allowing for complex functionality, increased reproducibility, and high throughput. However, LOC applications necessitate extensive sensor miniaturization to leverage these inherent advantages fully. Atom-sized quantum sensors are highly promising to bridge this gap and have enabled measurements of temperature, electric and magnetic fields on the nano- to microscale. Nevertheless, the technical complexity of both disciplines has so far impeded an uncompromising combination of LOC systems and quantum sensors. Here, we present a fully integrated microfluidic platform for solid-state spin quantum sensors, such as the nitrogen-vacancy (NV) center in diamond. Our platform fulfills all technical requirements, such as fast spin manipulation, enabling full quantum sensing capabilities, biocompatibility, and easy adaptability to arbitrary channel and chip geometries. To illustrate the vast potential of quantum sensors in LOC systems, we demonstrate various NV center-based sensing modalities for chemical analysis in our microfluidic platform, ranging from paramagnetic ion detection to high-resolution microscale NV-NMR. Consequently, our work opens the door for novel chemical analysis capabilities within LOC devices with applications in electrochemistry, high throughput reaction screening, bioanalytics, organ-on-a-chip, or single-cell studies.
\end{abstract}

\textbf{Keywords:} quantum sensing, nitrogen-vacancy (NV) center, magnetic resonance spectroscopy, nano- and microscale NMR spectroscopy, sensors, microfluidics, lab-on-a-chip, micro total analysis system

\begin{multicols}{2}
    \section{Introduction.}
\begin{figure*}[tbh]
    \centering
    \includegraphics[width=\linewidth]{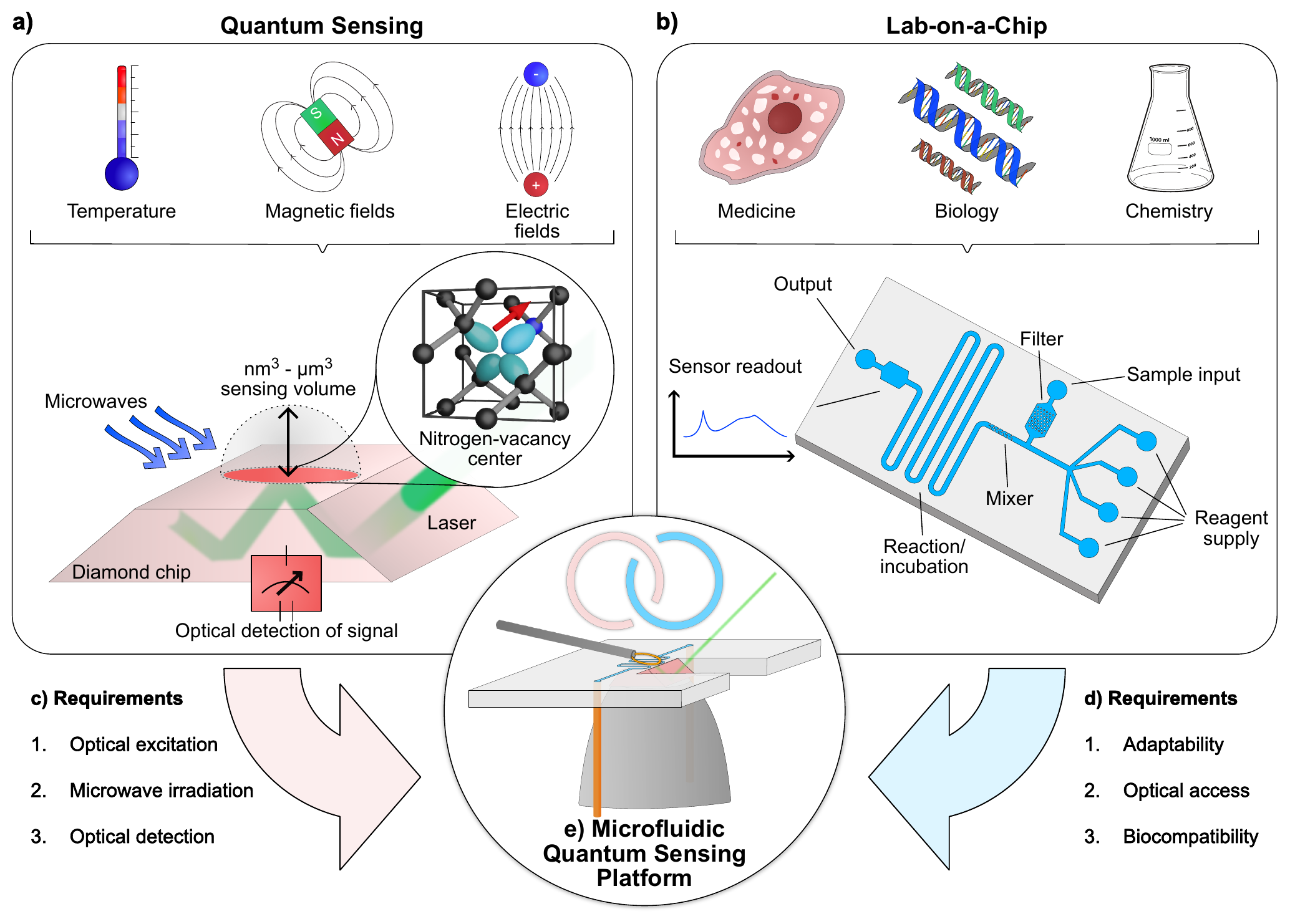}
    \caption{\textbf{Microfluidic quantum sensing platform.} 
    \textbf{a)} Quantum sensing of numerous physical quantities, \textit{e.g.}, temperature, electric-, and magnetic fields using nitrogen-vacancy (NV) centers in diamond. 
    \textbf{b)} Schematic of an example lab-on-a-chip (LOC) with various microfluidic operations for applications in, \textit{e.g.}, medicine, biology, or chemistry. 
    \textbf{c)} A list of experimental requirements posed by quantum sensing which must be incorporated into a generalized microfluidic sensing platform. 
    \textbf{d)} Requirements for the microfluidic quantum sensing platform to qualify for broad LOC applications.
    \textbf{e)} Our solution is a fully integrated microfluidic quantum sensing platform incorporating all criteria of quantum sensing and lab-on-a-chip applications.}
    \label{Figure_1}
    \end{figure*}

Lab-on-a-Chip (LOC) applications significantly impact physical and life sciences, hinging on the extensive system miniaturization of microfluidics, where benefits range from well-defined mass and heat transport to extensive process parallelization~\cite{Whitesides2006}. Additionally, microfluidics can optimize conventional chemical or biological processes~\cite{Mark2010-hy}, enabling applications in single-cell biology~\cite{Mehling2014-mw}, drug discovery~\cite{Dittrich2006-vf}, organ-on-a-chip research~\cite{Low2021-sh,Leung2022}, or electrochemistry~\cite{Jaugstetter2022}. However, analytics and process control in LOC platforms are still exceedingly challenging, requiring extensive sensor miniaturization and the integration of multiple sensor types~\cite{Kuswandi2007-vu,Livak-Dahl2011-ty,Jaugstetter2022}.\\
During the last decade, parallel to the developments in LOC, solid-state spin systems have emerged as promising atom-sized quantum sensors for nano- and microscale length scales (see Fig.~\ref{Figure_1}\textcolor{blue}{a})~\cite{Balasubramanian2008, Maze2008}. In particular, the nitrogen-vacancy (NV, see Fig.~\ref{Figure_1}\textcolor{blue}{a}~inset) center in diamond had an immense impact as a fully functional qubit at room temperature~\cite{Doherty2013-ue}. The working principle hinges on the strong interaction of (spin) qubits with their environment. While most quantum technologies aim to isolate their qubits as much as possible from their environment to limit interactions, quantum sensing harnesses these interactions for precision measurements of pressure, strain, temperature, electric or magnetic fields (see Fig.~\ref{Figure_1}\textcolor{blue}{a})~\cite{Degen2017_review,Schirhagl2014-hi}. Further, these physical quantities can often be measured multiplexed, allowing for parallel measurement of, \textit{e.g.}, temperature and magnetic fields~\cite{Shim2022-dv}. In particular, the magnetic sensing capabilities have been utilized extensively, \textit{e.g.}, to measure single-neuron action potential from live organisms~\cite{Barry2016}, high-resolution nuclear magnetic resonance (NMR) spectroscopy from single-cell volumes~\cite{Glenn2018-hp,Smits2019-kv,Bucher2020-dp,Arunkumar2021-gc}, or magnetic imaging of bacteria~\cite{LeSage2013}.
The main advantage stems from the fact, that the atom-sized sensors can be brought close to the diamond's surface (nano- to micrometers) and, thus, in contact with the sample which typically sets the sensing volume on a similar length scale (see Fig.~\ref{Figure_1}\textcolor{blue}{a})~\cite{Mamin2013,Staudacher2013,Bruckmaier2021,Glenn2018-hp}. However, most state-of-the-art experiments have not utilized the NV center's capabilities for nano- and microscale sensing to their fullest extent, still requiring milliliter sample volumes~\cite{Glenn2018-hp, Smits2019-kv,Bucher2020-dp,Arunkumar2021-gc}. Some early implementations have used polydimethylsiloxane (PDMS)-based~\cite{Ziem2013-mo,Steinert2013-gk} or tape-based~\cite{Smits2019-kv} microfluidic channels, with shortcomings in adaptability, biocompatibility, or limited quantum sensing capabilities. Thus, implementing NV center-based quantum sensing into a fully functional, unified LOC platform is a long-standing goal.\\
Here, we demonstrate a generalized microfluidic quantum sensing platform that incorporates the broad challenges posed by NV center-based quantum sensing and LOC applications. As example applications, we have chosen one of the most promising but challenging applications of NV centers in microfluidics: magnetic resonance spectroscopy~\cite{Allert2022-jl}, enabling non-invasive and quantitative analysis for chemical and biochemical applications on the nano- and microscale, complementing the existing NMR spectroscopy capabilities within microfluidics~\cite{Mompen2018, Barker2022, Eills2022}. Importantly, our platform promotes quantum sensing for other applications as well, such as temperature measurements~\cite{Neumann2013,Shim2022-dv}, bioassays~\cite{Glenn2015,Kayci2021}, biomagnetism~\cite{LeSage2013}, or velocimetry~\cite{Cohen2020-lu}, making this technology accessible to numerous research communities.
    \begin{figure*}[t]
    \centering
    \includegraphics[width=0.99\linewidth]{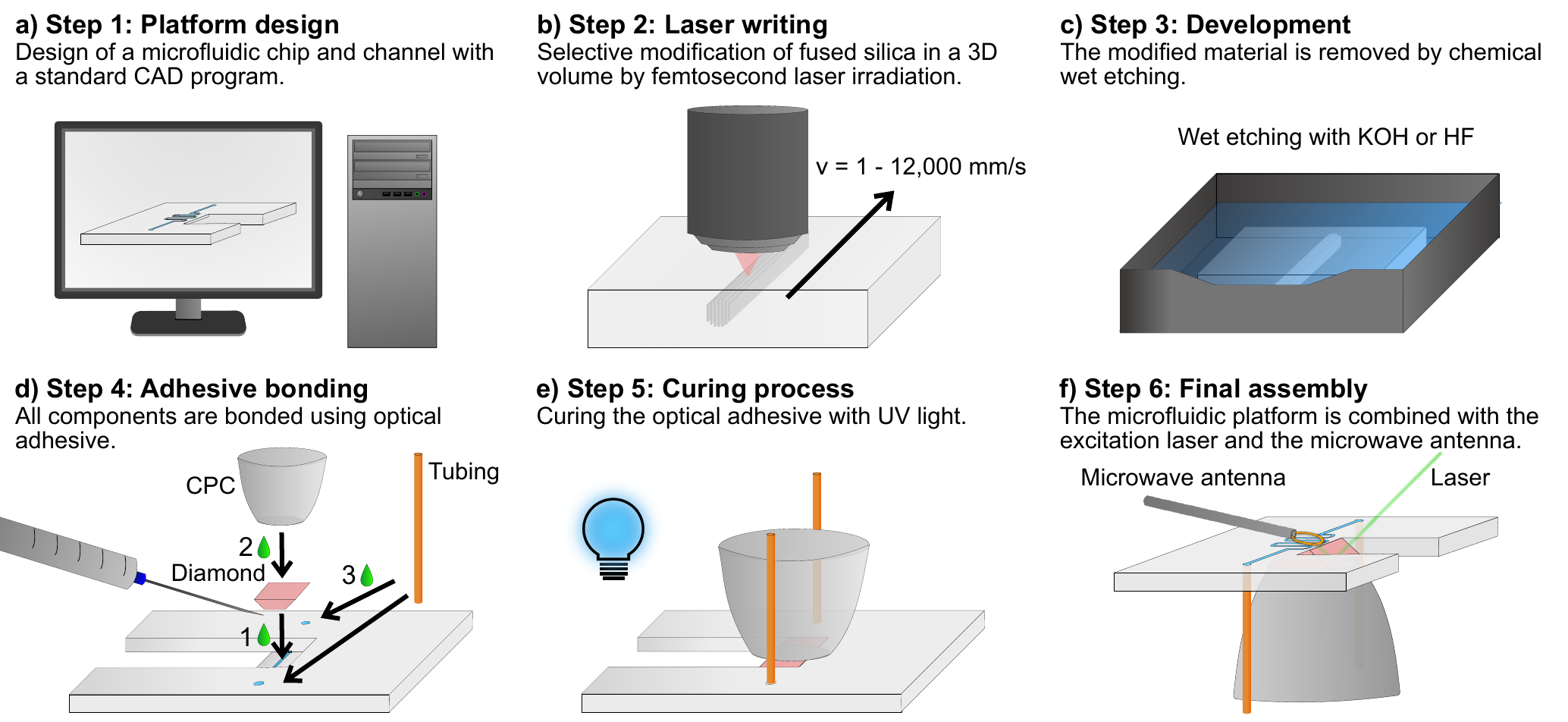}
    \caption{\textbf{Microfluidic chip manufacturing and assembly of the microfluidic quantum sensing platform.} \textbf{a)} The microfluidic chip is purposefully designed to match the required experimental conditions, \textit{e.g.}, diamond shape, light path and microfluidic operation, using commercial computer-aided design (CAD) software. \textbf{b)} The microfluidic chips are manufactured from optically polished fused silica, into which the three-dimensional chip structure is written using ultrashort infrared laser pulses. \textbf{c)} The chips are chemically wet-etched in an ultrasonic bath for several hours and up to days to expose the 3D microfluidic structure. \textbf{d)} The core of the microfluidic quantum sensing platform is assembled by first bonding of the NV-diamond in the microfluidic chip using optical adhesive; secondly, the compound parabolic concentrator (CPC) is bonded to the diamond, and lastly, the inlet and outlet tubing is bonded inside the microfluidic chip. \textbf{e)} The optical adhesive is subsequently cured with UV light. \textbf{f)} The final microfluidic quantum sensing platform is assembled and consists of the microfluidic chip, the NV-diamond, the CPC, tubing, microwave antenna, and the excitation laser.}
    \label{Figure_2}
    \end{figure*}
    \section{Microfluidic quantum sensing platform.}
    \subsection{Platform criteria.}\label{SubSec:MF_Critira}
A versatile microfluidic quantum sensing platform must fulfill various criteria posed by NV center-based quantum sensing and LOC applications, as illustrated in Fig.~\ref{Figure_1}\textcolor{blue}{c}~\&~\ref{Figure_1}\textcolor{blue}{d}. For quantum sensing with solid state spin systems, such as NV centers, the following technical equipment must be incorporated~\cite{Degen2017_review,Allert2022-jl}:
    \begin{enumerate}[label=\textbf{\arabic*}.]
        \item \textbf{Optical excitation.} The NV center's spin-state is initialized by laser excitation.
        \item \textbf{Microwave irradiation.} Microwave pulses are used for coherent spin-state manipulation in quantum sensing protocols. 
        \item \textbf{Fluorescence detection.} The measurement signal is read out by the NV center's spin-state dependent fluorescence.
    \end{enumerate}
Thus, these quantum sensors pose unique experimental requirements, such as high-power lasers (up to 1 - 10~\si{mW/\micro\meter^2} at $\sim$~532~nm~\cite{Glenn2018-hp,Levine2019-ov}) for spin-state initialization, efficient Gigahertz microwave delivery (up to a hundred watts) for spin-state control, and optical access for read-out (see Fig.~\ref{Figure_1}\textcolor{blue}{c}).
Furthermore, LOC applications pose requirements to the platform (see Fig.~\ref{Figure_1}\textcolor{blue}{d}), where we target the following:
\begin{enumerate}[label=\textbf{\arabic*}.]
    \item \textbf{Adaptability.} The microfluidic chip must be readily adaptable to facilitate various microfluidic operations (\textit{e.g.}, mixers) and to the experimental requirements for quantum sensing, such as diamond geometry, space constraints due to fluorescence light collection, or the laser excitation pathway.
    \item \textbf{Optical transparency.} The microfluidic chip must be optically transparent for microscopy and process control.
    \item \textbf{Biocompatibility.} Since many LOC applications study cells, organoids, or tissue, the biocompatibility of the platform is vital.
\end{enumerate}
It is necessary to design and manufacture the microfluidic quantum sensing platform as adaptable and biocompatible as possible to access the entire LOC application space. Furthermore, optical access is crucial for microscopy or additional spectroscopy methods. However, challenges arise from overlapping criteria of both fields, such as high-laser powers for quantum state initialization vs. biocompatibility or optical and microwave access vs. robustness of the microfluidic chip, which seem mutually exclusive. In this work, we overcome these obstacles and converge the presented criteria by a comprehensive engineering approach into an integrated microfluidic quantum sensing platform with broad applicability. These results are achieved by a unique combination of highly adaptable glass microfluidics, which is fully compatible with an optimized quantum sensing platform (see Fig.~\ref{Figure_1}\textcolor{blue}{e}).
    \subsection{Fabrication~\&~assembly process.} 
\begin{figure}[H]
    \centering
    \includegraphics[width=\linewidth]{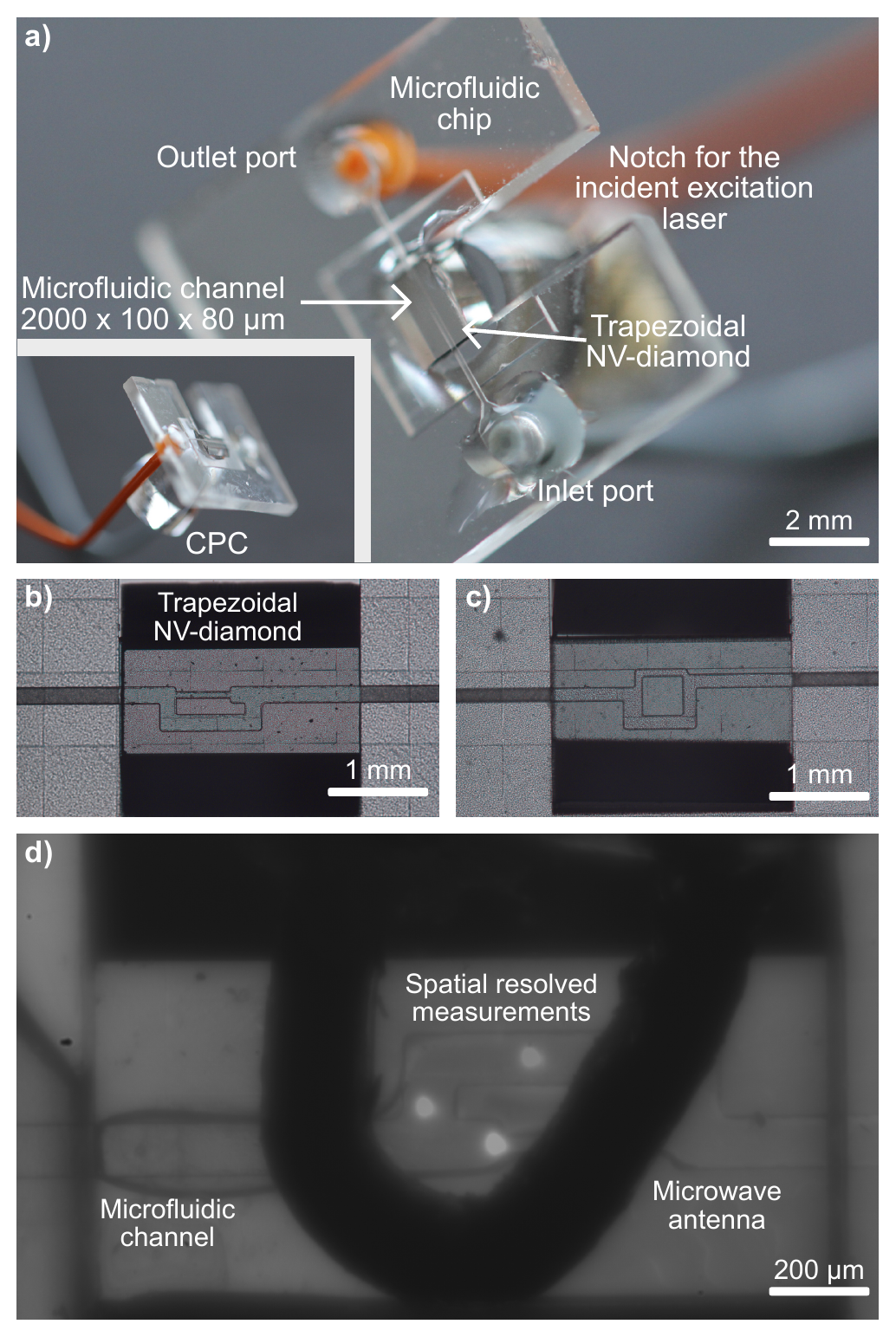}
    \caption{\textbf{Photographs of the integrated microfluidic quantum sensing platform.} \textbf{a)} Photograph of the microfluidic quantum sensing platform for a 100~\si{\micro\meter} thick and 80~\si{\micro\meter} high straight channel running the length of the 2.0 x 2.0~mm trapezoidal NV-diamond with 45\degree~slopes. The NV center-doped top side 2.0 x 1.0~mm acts as a sensor and forms the bottom of the channel. \textbf{b) \& c)} Example assembly with arbitrary channel geometries and dimensions demonstrate the platform's versatility. The angle polished sides of the diamond appear in dark. \textbf{d)} \textit{In operando} bright-field microscopy image of the complete assembly inside a microscale NV-NMR experiment: The microfluidic channel with varying dimensions runs across a 2.0 x 1.0~mm trapezoidal diamond. The coaxial microwave antenna appears as a dark loop positioned on the microfluidic chip. The active sensing area, \textit{i.e.}, the excitation area, is seen by the NV center's fluorescence (three sequential bright spots). Moving the excitation area enables scanning quantum sensing microscopy over the diamond's surface. Three overlayed pictures illustrate this imaging capability, demonstrating that measurements can be spatially resolved and addressed individually within the entire microfluidic structure.}
    \label{Figure_3}
    \end{figure}

We employ selective laser etching (SLE) to produce the microfluidic platform from quartz glass, \textit{i.e.}, synthetic fused silica~\cite{Gottmann2017-ok,Butkut2021}, a well-established technique for laser-assisted micromachining of three-dimensional devices in transparent dielectrics. Compared to other techniques, SLE based-glass chips allow for arbitrary shapes and three-dimensional designs, fulfilling complex geometry requirements for quantum sensing integration (see Fig.~\ref{Figure_2}\textcolor{blue}{a}~\&~\ref{Figure_3}\textcolor{blue}{a}). Notably, SLE enables the fabrication of monolithic internal channels, \textit{i.e.}, channels inside the 3D substrate, crucial for integrating NV-diamonds where the surface area for bonding is typically minimal. SLE utilizes ultrashort femto- to picosecond infrared (IR) laser pulses to locally modify the substrate, forming nanograting-like structures at the focus point (see Fig.~\ref{Figure_2}\textcolor{blue}{b})~\cite{Hnatovsky2005}. Three-dimensional spatial selectivity is achieved due to the highly nonlinear nature of the light-matter interaction in the focal volume~\cite{Hnatovsky2005}. The modified glass is subsequently wet-etched with either potassium hydroxide or hydrogen fluoride solution (see Fig.~\ref{Figure_2}\textcolor{blue}{c}) with significantly increased etching rates compared to non-illuminated sections, thus, exposing the negative 3D image~\cite{Butkut2021, Gottmann2017-ok}. The etch-rate increase of the laser-modified material can be more than 1,400:1 in quartz silica~\cite{Gottmann2017-ok}, giving rise to high selectivity and, thus, high spatial resolution ($\sim$~1~\si{\micro\meter} in XY and $\sim$~2~\si{\micro\meter} in the Z-direction,~\cite{Butkut2021}). The main advantage is that 3D internal channels can be fabricated as long as the total distance to the next opening is $\sim$~10~mm~\cite{Gottmann2017-ok} to ensure sufficient exchange of etching medium.
After the chip fabrication, the diamond is glued into the microfluidic chip using an optical adhesive, forming the channel bottom and, thus, sealing the chip (see Fig.~\ref{Figure_2}\textcolor{blue}{d}). Next, a compound parabolic concentrator (CPC) for enhanced fluorescence light collection is bonded to the backside of the diamond~\cite{Wolf2015_subpT}. Then, the in- and output of the microfluidic chip are connected to the microfluidic system using commercial high-performance liquid chromatography (HPLC) tubing and additional optical adhesive to secure the tubing inside the chip. Finally, the combined assembly is cured under ultra-violet (UV) illumination (see Fig.~\ref{Figure_2}\textcolor{blue}{e}) before the laser and microwave delivery can be incorporated (see Fig.~\ref{Figure_2}\textcolor{blue}{f}~\&~\ref{Figure_3}\textcolor{blue}{a}).
We want to emphasize that the microfluidic assembly can quickly be disassembled and cleaned by dissolving the optical adhesive either with hot sulfuric acid or chlorinated solvents such as dichloromethane. Therefore, the microfluidic platform can be easily reassembled and reused after cleaning.
    \subsection{Results.}
\begin{figure}[H]
    \centering
    \includegraphics[width=\linewidth]{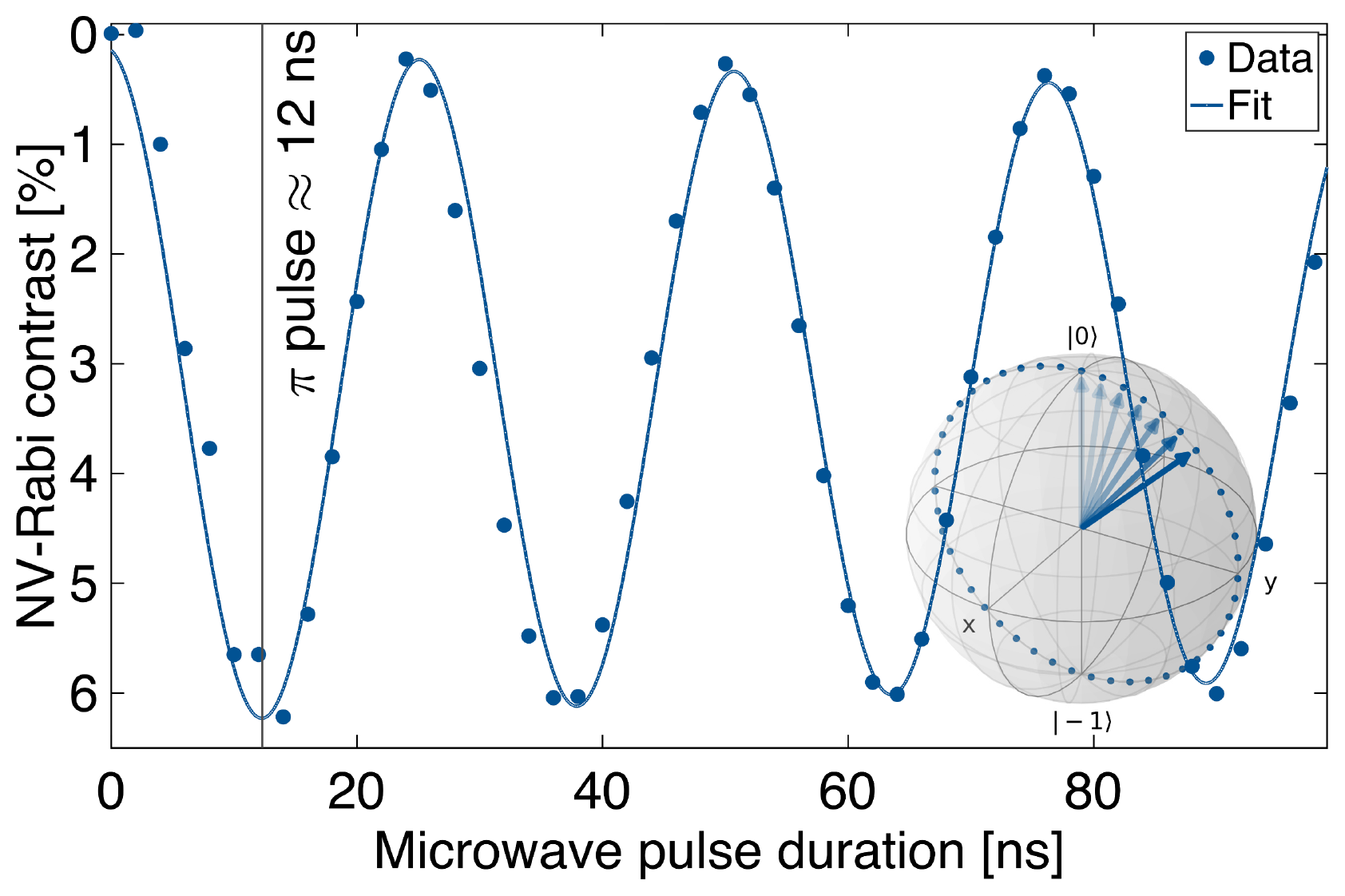}
    \caption{\textbf{NV center spin control inside the microfluidic chip.} Measured Rabi experiment where the $\ket{0} \to \ket{-1}$ transition at $\sim$~1.95~GHz was resonantly driven using a coaxial microwave antenna placed on the microfluidic assembly. Rabi frequencies above $\sim$~40~MHz, equivalent to a $\pi$ pulse duration of $\sim$~12~\si{\nano\second}, are achieved, enabling a broad range of quantum sensing schemes. The experimental data is shown as blue dots, which is fitted using a decaying sinusoidal function. Inset: Bloch sphere representation of the Rabi experiment.}
    \label{Figure_4}
    \end{figure}

In order to fulfill the challenging requirements, the engineering process for the microfluidic chip started with an optimized NV-experiment~\cite{Glenn2018-hp,Bucher2019-yu,Liu2022-pj} and their respective applications. We then fabricated numerous microfluidic chips using the described SLE fabrication process and validated the platform based on the target criteria established in Section~\ref{SubSec:MF_Critira}. As a result, several platform generations have been produced and validated with an step-wise increase in complexity, implementing more criteria with every generation. The concluding fully assembled microfluidic quantum sensing platform is depicted in Fig.~\ref{Figure_3}\textcolor{blue}{a}, consisting of the microfluidic chip with a 2,000 (length) x 100 (width) x 80 (height)~\si{\micro\meter} straight channel running the length of the 2.0 x 2.0~mm trapezoidal NV-diamond. The channel runs inside the microfluidic chip from the in- and outlet port to the diamond, which then forms one of the channel walls. The total volume of the microfluidic channel is $\sim$~600~nL with an active sensing volume of, \textit{e.g.}, $\sim$~130~pL, defined by the area of optically excited NV centers (d~$\approx$ 45~\si{\micro\meter}, see Fig.~\ref{Figure_3}\textcolor{blue}{d}) and the height of the microfluidic channel (h~$\approx$~80~\si{\micro\meter}). More complex microfluidic structures are shown in Fig.~\ref{Figure_3}\textcolor{blue}{b}~\&~\ref{Figure_3}\textcolor{blue}{c} (see SI Section~\ref{Figure_SI2} for technical drawings).\\
\textbf{Validation of quantum sensing criteria.}
Firstly, the NV center must be optically initialized for quantum sensing with laser powers up to 1 - 10~\si{mW/\micro\meter^2} at $\sim$~532~nm~\cite{Glenn2018-hp,Levine2019-ov}. These high power intensities severely limit the application of NV center-based quantum sensing in biological applications, as direct illumination of cells with these laser powers can lead to cell death within tens of seconds~\cite{Wldchen2015}. This can be mitigated by keeping the light within the diamond chip by a total internal reflection (TIR) geometry of the NV center's excitation laser path can limit the sample's exposure to evanescent light, significantly reducing photodegradation~\cite{Glenn2018-hp}. Additionally, the TIR geometry could significantly increase the laser power density in the excited volume by passing twice through the NV-doped layer. This effect occurs if the excited area (laser beam diameter) is on the same order of the NV-layer thickness. Typically, micrometer-thick NV ensembles profit from this geometry. Consequently, we polished the sides of a 2.0 x 2.0 x 0.5~mm diamond chip at a 45\degree~angle and exposed one of the polished sides to the incident laser (see Fig.~\ref{Figure_3}\textcolor{blue}{a}), enabling a TIR geometry in the excitation pathway (see Fig.~\ref{Figure_1}\textcolor{blue}{a}). The adaptability and robustness of SLE process allows for designing a microfluidic chip with notches and delicate structures that do not interfere with the TIR laser path for NV center-based quantum sensing (see Fig.~\ref{Figure_1}\textcolor{blue}{a}). Furthermore, this geometry allows for straightforward manipulation of the excitation laser position, \textit{e.g.}, via a kinematic mirror, facilitating scanning quantum sensing microscopy over the whole diamond surface. Fig.~\ref{Figure_3}\textcolor{blue}{d} shows the excitation of three different sensing areas in three overlayed pictures.\\
Secondly, quantum sensing requires fast and efficient spin control~\cite{Degen2017_review} with microwaves. The contact between the microwave delivery structures and the chemical or biological samples should be avoided. For that reason, we place the microwave antenna in our platform on top of the microfluidic chip (see Fig.~\ref{Figure_2}\textcolor{blue}{f}~\&~\ref{Figure_3}\textcolor{blue}{d}). Crucially, the distance between the microwave antenna and the NV centers needs to be minimized as the microwave power falls off with distance. The fused silica's stiffness allows for thin channel ceilings ($\sim$~120~\si{\micro\meter}), reducing the distance between the diamond's surface and the microwave antenna to $\sim$~200~\si{\micro\meter} (including the microfluidic channel) without sacrificing stability of the microfluidic structures. Using a high-power microwave amplifier resonantly driving the NV center transition at $\sim$~1.95~GHz (see Methods section), Rabi frequencies above $\sim$~40~MHz equivalent to a $\pi$ pulse duration of $\sim$~12~ns can be achieved (see Fig.~\ref{Figure_4}), enabling full quantum sensing capabilities~\cite{Levine2019-ov,Allert2022-jl}. Importantly, our design allows for imaging of various physical quantities over a large field of view within the microfluidic structure (see Fig.~\ref{Figure_3}\textcolor{blue}{d}) - a unique characteristic solid-state quantum sensors. The employed microwave antenna, made from a semi-rigid coaxial cable, has the advantage that a broad frequency range can be excited compared to resonant structures. However, more complex microwave antennas from printed circuit boards or microwave resonators can be placed on the microfluidic chip for increased power delivery. We note that the presented microfluidic chip could be utilized as a glass substrate for sputtering microwave antennas, such as strip lines, omega antennas~\cite{Opaluch2021-iw}, or resonant structures~\cite{Mariani2020-nz} for further integration and miniaturization of the experiment in the future.\\
Lastly, the NV spin-state is read-out via its fluorescence intensity to translate the quantum measurement into a detectable signal. Several methods for fluorescence collection have been established, most prominently via microscope objectives or optical elements such as solid-immersion lenses (SILs) or CPCs. Our platform facilitates all three of these methods; however, we choose to optimize our assembly for a CPC as it has been established as one of the most efficient light collection strategies for NV ensembles, with up to 65\% collection efficiency~\cite{Wolf2015_subpT}. Fig.~\ref{Figure_3}\textcolor{blue}{a} depicts the fully assembled system, including the CPC. Our presented platform, thus, meets all three fundamental criteria posed by NV center-based quantum sensing: It incorporates TIR excitation for efficient NV center initialization without compromising biocompatibility, fast NV spin manipulation, and state-of-the-art light collection strategies. We would like to note, that our platform can easily be adapted for single NV center experiments using confocal microscopes (see SI Section~\ref{SI_Chipdesign}).\\
\textbf{Validation of LOC criteria.}
Utilizing SLE fabrication and, in extension, glass microfluidics has several advantages: one of the main benefits is the platform's adaptability. The microfluidic chip can be tailor-made to the requirements of quantum sensing, \textit{e.g.}, delicate structures due to diamond shape or microwave delivery, and LOC applications by complex microfluidic channel designs. As a second example, with less stringent optical requirements, we designed a different chip geometry for a standard 2.0 x 2.0 x 0.5~mm cuboid diamond, which can be found in SI Section~\ref{SI_Chipdesign}. The channel width in fused silica using SLE can be as low as $\sim$~10~\si{\micro\meter}, and the channel height $\sim$~20~\si{\micro\meter}; though, in our experiments, we typically employ channels between 40 - 100~\si{\micro\meter} in width and 30 - 80~\si{\micro\meter} in height which reduces the internal pressure and represents similar length scales as microscale NV center-based sensing (see Fig.~\ref{Figure_3}\textcolor{blue}{a}). Especially, the material's stiffness allows for a wide range of channel geometries and aspect ratios, see Fig.~\ref{Figure_3}\textcolor{blue}{b}~\&~\ref{Figure_3}\textcolor{blue}{c} for examples incorporating fluidic bypass structures for measurements in reduced volumes. These complex designs can even be implemented and sealed on small diamond surfaces (2.0 x 1.0~mm) without difficulties.\\
The combination of design freedom in channel geometry with the high spatial resolution of SLE allows for the generation of various microfluidic operations (\textit{e.g.}, mixers, cell traps, or picoinjectors) directly on top of the diamond surface, \textit{i.e.}, the sensing area, which permits non-invasive studying of processes taking place at the point of interest. Furthermore, quartz glass is optically transparent, chemical resistant, and biocompatible, encouraging long-term studies in applications such as organ-on-a-chip studies~\cite{Low2021-sh,Leung2022}. Glass microfluidics has further been demonstrated with femtoliter reaction chambers using electron beam lithography for further miniaturization~\cite{Kazoe2019}.\\
    \section{Example applications.}\label{Sec:MF_MagRes}
\begin{figure*}[phtb]
    \centering
    \includegraphics[width=\linewidth]{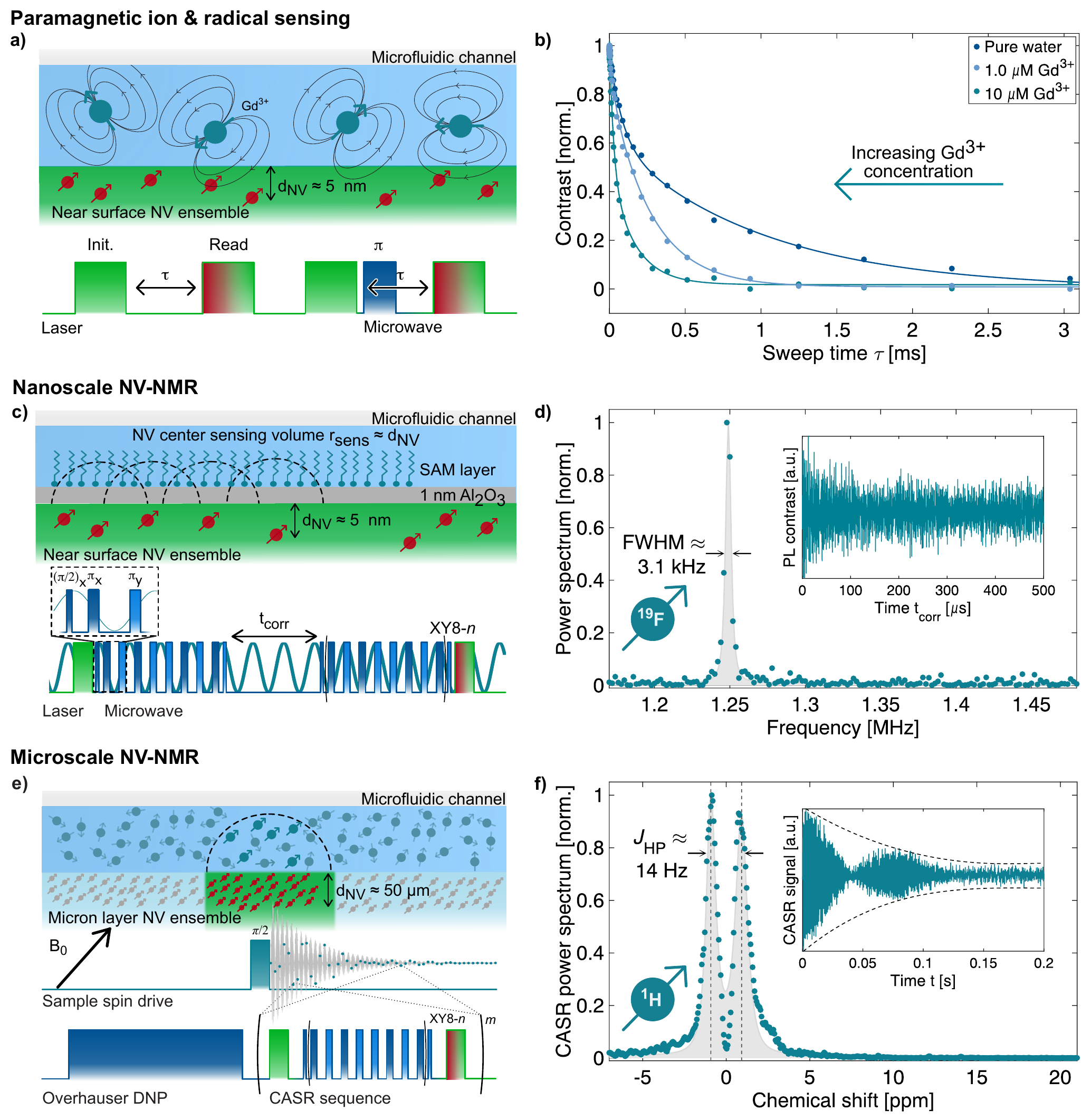}
    \caption{\textbf{Magnetic resonance measurements inside the microfluidic quantum sensing platform.} 
    \textbf{a)} Schematic of paramagnetic ion sensing in microfluidics measuring the T$_1$ relaxation time of near-surface NV centers.
    \textbf{b)} NV-T$_1$ relaxation curves of pure water, 1.0~\si{\micro M}, and 10~\si{\micro M} Gd$^{3+}$ solutions. The NV-T$_1$ relaxation rate increases with increasing Gd$^{3+}$ concentration, which demonstrates the high sensitivity of the NV center quantum sensors for paramagnetic ions and radicals. 
    \textbf{c)} Schematic of nanoscale NV-NMR probing covalently bound fluorinated molecules an amorphous aluminum oxide (Al$_2$O$_3$) layer on the diamond surface utilizing correlation spectroscopy. 
    \textbf{d)} Power spectrum of the $^{19}$F NMR signal originating from the surface-tethered molecules within the microfluidic channel. Inset: time domain correlation signal of the $^{19}$F. 
    \textbf{e)} Schematic of high-resolution microscale NV-NMR detecting trimethyl phosphate (TMP) inside the microfluidic channel using a 50~\si{\micro\meter}-thick NV center-doped layer. The signal is detected with a CASR-type pulse sequence in combination with Overhauser hyperpolarization for signal enhancement.
    \textbf{f)} Power spectrum of the $^1$H-NMR signal of trimethyl phosphate (TMP) inside the microfluidic channel demonstrating high spectral resolution. Inset: time domain of the CASR-detected NMR signal.}
    \label{Figure_5}
    \end{figure*}

We opted for magnetic resonance applications to illustrate the NV center's unique capabilities to detect nuclear and electronic spins with unprecedented sensitivities~\cite{Staudacher2013,Lovchinsky2016}, ideal for LOC applications.\\
\textbf{Paramagnetic ion~\&~radical sensing.} The NV center can detect paramagnetic ions and radicals with high sensitivity by relaxometry~\cite{Steinert2013-gk,Ziem2013-mo}. In this sensing scheme, magnetic noise caused by unpaired electronic spins leads to an increase in the NV center relaxation, probed by the spin-state-depended fluorescence. NV-T$_1$ relaxometry has been successfully applied to study, \textit{i.e.}, the cellular response to antibiotics~\cite{Norouzi2022} and viral infections~\cite{Wu2022}, or to monitor chemical reactions \textit{in situ}~\cite{Simpson2017}. The measurement of T$_1$ relaxometry of paramagnetic species has also been demonstrated under physiological conditions~\cite{Ziem2013-mo,Steinert2013-gk}.\\
Here, we realized T$_1$ relaxometry sensing of paramagnetic ions inside the microfluidic quantum sensing platform using a near-surface NV center ensemble (see Materials and Methods). The NV-T$_1$ relaxometry protocol is depicted in Fig.~\ref{Figure_5}\textcolor{blue}{a}. The measurements are conducted by filling the microfluidic channel (straight channel see SI Section~\ref{SI_Chipdesign}) with either pure water as a reference sample or gadolinium ion solution (1.0~\si{\micro M} and 10~\si{\micro M}). Fig.~\ref{Figure_5}\textcolor{blue}{b} depicts the concentration-dependent T$_1$ relaxation, which demonstrates the high sensitivity of our microfluidic quantum sensing platform. The microfluidic chip allows for highly controlled experimental conditions, which were impossible before~\cite{freire-moschovitis_diamag-_2022}. This experiment demonstrates that the microfluidic quantum sensing platform can be applied to detect radicals or paramagnetic ions, which sets the route for electrochemical or biomedical sensing.\\
\textbf{Nanoscale NV-NMR.} NMR spectroscopy at the nanoscale (a few nanometer diameter detection volumes) using NV centers has demonstrated NMR spectra from single proteins covalently bound to the diamond's surface~\cite{Lovchinsky2016}, 2D materials~\cite{Lovchinsky2017}, or probing the formation of a self-assembled monolayer (SAM) in real time under chemically relevant conditions~\cite{Liu2022-pj}.\\
We detected NMR signals from surface tethered molecules (12-pentafluorophenoxydodecylphosphonic acid, PFPDPA) within a microfluidic chip with a straight channel (see SI Section~\ref{SI_Chipdesign}). We employ correlation spectroscopy (see Fig.~\ref{Figure_5}\textcolor{blue}{c}) to detect the $^{19}$F-NMR signal. The $^{19}$F correlation spectroscopy data results in a time domain NMR signal (see Fig.~\ref{Figure_5}\textcolor{blue}{d} inset). The Fourier transform of these data results in the $^{19}$F-NMR power spectrum with a resonance at 1.25~MHz (see Fig.~\ref{Figure_5}\textcolor{blue}{d}). These results demonstrate that the established method of nanoscale NV-NMR can readily be interfaced with our microfluidic platform, opening up the route to a broad application space in materials science studying solid-liquid interfaces. Furthermore, recent studies have utilized Al$_2$O$_3$-modified diamond surfaces to create a biocompatible surface architecture~\cite{Xie2022-yh}. The combination of microfluidics, nanoscale NV-NMR, and biocompatible surface modifications paths the way to label-free, high-throughput screening or analysis of biomolecules.\\
\textbf{Microscale NV-NMR.} While nanoscale NV-NMR uses near-surface NV centers, microscale NV-NMR typically uses micrometer-thick diamond layers homogeneously-doped with NV centers~\cite{Allert2022-jl}. These micrometer-thick layers allow the detection of NMR signals in picoliter volumes with high spectral resolution in contrast to the nanoscale experiments (see Fig.~\ref{Figure_5}\textcolor{blue}{d}~\&~\ref{Figure_5}\textcolor{blue}{f})~\cite{Allert2022-jl}, a highly promising technology for LOC. However, due to the lack of suitable microfluidic platforms, previous studies~\cite{Glenn2018-hp,Smits2019-kv,Bucher2020-dp,Arunkumar2021-gc} still utilized milliliters of sample volume, omitting the benefits of NV-NMR. Here, we demonstrate microscale NV-NMR with at least two orders of magnitude less sample volume in the microfluidic chip than in previous studies~\cite{Smits2019-kv}. We utilized a diamond sensor with a 50~\si{\micro\meter} NV-doped layer, initialized via a TIR excitation geometry with a laser spot diameter of $\sim$~45~\si{\micro\meter}, resulting in a detection volume of $\sim$~130~pL. The NV-doped diamond layer thickness was specifically developed and produced to compliment the 100 x 80~\si{\micro\meter} microfluidic channel, shown in Fig.~\ref{Figure_3}\textcolor{blue}{a} (see SI Section~\ref{SI_Sec_layeroptimization})~\cite{Bruckmaier2021}. The measurement procedure follows Bucher \textit{et al.}~\cite{Bucher2020-dp} using the coherently averaged synchronized read-out (CASR) protocol~\cite{Glenn2018-hp} combined with Overhauser DNP for signal enhancement (see Fig.~\ref{Figure_5}\textcolor{blue}{e}). As an example, the $^{1}$H-NMR spectrum of trimethyl phosphate (TMP) was acquired at $\sim$~180~mT.
Fig.~\ref{Figure_5}\textcolor{blue}{f} depicts the time domain of the CASR signal (inset) and the power spectrum. The linewidth is $\sim$~5~Hz ($\sim$ 0.7 ppm), and the \textit{J}-coupling between $^{31}$P and $^{1}$H of 14~Hz is observable. Currently, the spectral resolution is limited by the homogeneity of the applied magnetic field, which could be improved by, \textit{e.g.}, incorporating shim structures into the microfluidic chip~\cite{Hale2018,vanMeerten2018}. In addition, the magnetic field strength for NV-NMR spectroscopy is currently limited by the achievable Rabi frequency of the NV centers~\cite{Allert2022-jl}, which can be improved by utilizing more efficient microwave antennas, \textit{e.g.}, microwave resonators. Furthermore, we observe a channel size-dependent NMR signal, indicating that the sensing volume matches the microfluidic length scales~\cite{Bruckmaier2021}.\\
We demonstrated that the microfluidic quantum sensing platform can be used for high-resolution, microscale NV-NMR, opening up the route toward promising applications such as \textit{in vitro} single-cell or organoid studies, NMR-based microdroplet sorting, and process control in microfluidics.
    \section{Conclusion.}
In conclusion, we developed, fabricated, and applied a novel technology platform for combining Lab-on-a-chip applications with solid-state quantum sensors. Our microfluidic quantum sensing platform facilitates various diamond and microfluidic channel layouts and enables fast NV center spin control - vital for quantum sensing applications. Furthermore, our platform is biocompatible by utilizing a TIR geometry in its excitation pathway, limiting the photodegradation of chemical or biological samples. Lastly, our platform allows for optical access for the NV center's fluorescence read-out and simultaneous optical microscopy.\\
The combination of the two enabling technologies, LOC and quantum sensors, in a single, versatile, and adaptable platform will have a distinct impact on numerous research fields in the physical and life sciences. The microfluidic quantum sensing platform will open up novel imaging and sensing capabilities, such as label-free and non-invasive chemical analysis on a chip~\cite{Allert2022-jl}. We are convinced that both research fields of quantum sensing and LOC will benefit significantly from this novel analytical tool with applications in electrochemistry~\cite{Jaugstetter2022}, drug~\cite{Dittrich2006-vf} and catalysis screening, bioanalytics~\cite{Leung2022}, or single-cell biology~\cite{Mehling2014-mw}.
    \section*{Materials and Methods}
\textbf{Microfluidic quantum sensing platform assembly.} The microfluidic chip was designed with a computer-aided design (CAD) software (Fusion360, Autodesk, Mill Valley, United States) and manufactured by LightFab GmbH (Aachen, Germany) using the LightFab 3D Printer. The process started from a 1.0~mm thick optically polished fused silica chip, in which the three-dimensional structure was written into using ultrashort (500 - 2,000~fs with an average power of $\sim$~200 - 2,000~mW at 250 - 3,000~kHz repetition rate) IR laser pulses at the wavelength of 1030~nm. Subsequently, chemical wet-etching in 8~mol/L KOH at 85~\degree C in an ultrasonic bath for several hours and up to days exposed the 3D microfluidic structure. Then the NV-diamond was glued into the microfluidic chip using optical adhesive (NOA68, Norland Products, Jamesburg, United States), which was cured for >~12~h at 50~\degree C under UV illumination (Form Cure, Formlabs Inc., Somerville, United States). Similarly, a CPC was glued to the backside of the diamond. The CPC was designed in-house and fabricated by Süd-Optik Schirmer GmbH (Kaufbeuren, Germany). The liquid samples were injected through polyether ether ketone (PEEK) tubing (\textit{e.g.}, SGEA1301005001-5F, VWR, Randor, United States).\\
\textbf{Preparation of near-surface NV centers in diamond.}
The near surface NV-diamond sample was produced according to Liu \textit{et al.}~\cite{Liu2022-pj}. In short, an electronic grade single crystal diamond (2.0 x 2.0 x 0.5~mm) with a bulk nitrogen concentration of <~5~ppb (Element Six Technologies Limited, Didcot, United Kingdom) was implanted with $^{15}$N at an energy of 2.5~keV with an off-axis tilt of 7\degree~and a total ion fluence of $2*10^{12}$~\si{1/cm^2} by II-VI Incorporated (San Jose, United States). The implanted diamond was subsequently annealed under vacuum in a home-built oven for >~32~h at incremental temperature steps up to $\sim$~800~\degree C. In addition, the diamond was cleaned before and after every production step with a triacid cleaning protocol involving equal volumetric quantities of sulfuric, nitric, and perchloric acid at $\sim$~200~\degree C for $\sim$~2~h.\\
\textbf{Preparation of microscale NV center-doped layers in diamond.} The NV center-doped diamond micron-layer ($\sim$~50~\si{\micro\meter}, $\sim$~19~ppm substitutional nitrogen (P1 centers)) was homoepitaxially grown on electronic grade single crystal diamonds (2.0 x 2.0 x 0.5~mm) with a bulk nitrogen concentration of <~5~ppb (Element Six Technologies Limited, Didcot, United Kingdom) by chemical vapor deposition (CVD) at the Fraunhofer Institute for Applied Solid State Physics (Freiburg, Germany)~\cite{Schtzle2022}. The diamond sample was subsequently irradiated with electrons and high-temperature annealed under ultrahigh vacuum, yielding a dense NV ensemble layer~\cite{Glenn2018-hp}. Subsequently, two sides of the diamonds were polished at a 45\degree~angle yielding trapezoidal-shaped diamonds where the top face is 2.0 x 1.0~\si{mm} in size (MEDIDIA GmbH, Idar-Oberstein, Germany). The trapezoidal shape enabled internal reflection of the excitation laser.\\
\textbf{NV-Rabi experiment.} The Rabi measurements were conducted in an experiment setup previously described by Liu \textit{et al.}~\cite{Liu2022-pj}. The experiments were performed by resonantly driving the $\ket{0} \to \ket{-1}$ transition at $\sim$~1.95~GHz ($\sim$~33~mT) with a 100~W microwave amplifier (KU PA BB 070270-80 A-2.1.1, Kuhne electronic GmbH, Berg, Germany). The external magentic field was applied according to Bucher \textit{et al.}~\cite{Bucher2019-yu} using commercially available neodymium permanent magnets and optomechanics. Furthermore, as in all following NV-experiments, the external magnetic field was aligned along the <111> diamond lattice direction.\\
\textbf{NV-T$_1$ relaxometry.} The NV-T$_1$ measurements were conducted in an experiment setup previously described by Liu \textit{et al.}~\cite{Liu2022-pj}. The pulse sequence is depicted in Fig.~\ref{Figure_5}\textcolor{blue}{a}. The NV-T$_1$ relaxation experiments were performed at $\sim$~33~mT and the sweep time $\tau$ was increased in 51 logarithmically-spaced steps from 200~ns to 5.5~ms~\cite{freire-moschovitis_diamag-_2022}. Every data point was averaged over 5,000 single measurements. The T$_1$ relaxation data was normalized and fitted with a biexponential function~\cite{freire-moschovitis_diamag-_2022}. The electrolyte solutions were prepared using deionized water with a resistivity of 18.2~\si{\mega\ohm\centi\meter} (MilliporeSigma,  Burlington, United States) in which gadolinium(III) nitrate hydrate (11284, Alfa Aesar by Thermo Fisher Scientific, Haverhill,United States) was dissolved.\\
\textbf{Nanoscale NV-NMR.} The nanoscale NV-NMR experiment was performed in the experimental setup previously described in Liu \textit{et al.}~\cite{Liu2022-pj}. The pulse sequence is depicted in Fig.~\ref{Figure_5}\textcolor{blue}{c}. The correlation spectroscopy was performed at 31~mT using two dynamically decoupling (XY8-4) blocks. The time $t_{corr}$ between two XY8-4 blocks was swept starting from 2\si{\micro\second} to 502~\si{\micro\second} in 2,501 equally spaced steps. Each point was averaged over 10,000 times and the entire experiment was repeated 8 times. The time domain data was Fourier transformed (MATLAB R2022a, MathWorks, Natick, United States) and the power spectrum was fitted with a modified Lorentzian function (see grey area in Fig.~\ref{Figure_5}\textcolor{blue}{d})~\cite{Glenn2018-hp}. The surface tethered molecule sample was prepared according to Liu \textit{et al.}~\cite{Liu2022-pj}. In short: The Al$_2$O$_3$ layer was deposited with atomic layer deposition (ALD) performed on a Veeco Fiji G2 ALD system (Plainview, United States) for a final layer thickness of $\sim$~1~nm. The sample was activated by oxygen plasma and subsequently exposed to a 12-pentafluorophenoxydodecylphosphonic acid (PFPDPA) solution in ethanol (10~mM, 777188-500MG, Sigma-Aldrich, St. Louis, United States) for two days and rinsed with ethanol to remove physisorbed molecules before the measurement.\\
\textbf{Microscale NV-NMR.} The microscale NV-NMR experiment was performed in the experimental setup described in Bruckmaier \textit{et al.}~\cite{bruckmaier2022-diff}. The measurement was conducted at a magnetic field of $\sim$~180~mT, which was applied via a large bore superconducting magnet (3T-215-RT, Superconducting Systems INC., Billerica, United States). The pulse sequence is depicted in Fig.~\ref{Figure_5}\textcolor{blue}{e} and consisted of a 0.3~s microwave pulse for Overhauser DNP at the resonance frequency (4.9~GHz) of the free electron of Tempol, followed by a proton $\pi/2$ pulse at 7.6~MHz with a duration of $\sim$~46~\si{\micro\second}, and the CASR detection sequence~\cite{Glenn2018-hp}. The CASR detection sequence was based on a UDD12-55 subsequences for dynamic decoupling~\cite{Biercuk2009}, which is repeated 9,470 times within 1.0 second. The recorded data set was averaged 100 times. After Fourier transformation of the time domain data set, the power spectrum was fitted with a modified double Lorentzian function (see grey area in Fig.~\ref{Figure_5}\textcolor{blue}{f})~\cite{Glenn2018-hp}. The center frequency was defined as chemical shift $\delta = 0~\si{ppm}$ for visualization. The experiment was performed with pure trimethyl phosphate (TMP, 241024, Sigma-Aldrich, St. Louis, United States) and 20~mM Tempol (581500, Sigma-Aldrich, St. Louis, United States).
    \section*{Author Contributions}
R.D.A., F.B., N.R.N., and D.B.B. devised, designed, engineered, and validated the microfluidic chip. C.S. and M.H. fabricated the microfluidic chip. R.D.A., F.B., N.R.N., F.A.F.-M., K.S.L, and D.B.B. designed and constructed the NV-experiments. R.D.A., F.B., N.R.N., F.A.F.-M., and K.S.L. performed the measurements for magnetic resonance experiments and analyzed the data. R.D.A., P.S., and P.K. produced the NV-diamond for microscale NMR. R.D.A. and D.B.B. wrote and edited the manuscript with input from all coauthors.
    \section*{Declaration of Competing Interest}
R.D.A., F.B., N.R.N., F.A.F.-M., K.S.L., P.S., P.K., and D.B.B. declare that they have no known competing financial interests or personal relationships that could have appeared to influence the work reported in this article. The authors C.S. and M.H. are part of the company LightFab GmbH and have an interest in selling manufactured glass parts and the LightFab 3D Printer. Educating the constructers on the new possibilities provided by SLE may increase the number of manufactured glass parts sold. No funding sponsors are involved in LightFab GmbH.
    \section*{Acknowledgments}
This work was supported by the European Research Council (ERC) under the European Union’s Horizon 2020 research and innovation program (grant agreement No 948049) and by the Deutsche Forschungsgemeinschaft (DFG, German Research Foundation) – 412351169 within the Emmy Noether program. D.B.B. acknowledges support by the DFG under Germany’s Excellence Strategy—EXC 2089/1—390776260 and the EXC-2111 390814868. P.S. and P.K. acknowledge financial support by the Fraunhofer Lighthouse Project Quantum Magnetometry (QMag).
    \section*{Data and materials availability}
All data needed to evaluate the conclusions in the paper are present in the paper and/or the supplementary information. Additional data related to this paper may be requested from the authors. All correspondence and request for materials should be addressed to D.B.B. (dominik.bucher@tum.de).
    \section*{References}\printbibliography[heading=none]
\end{multicols}
    \newpage
    \appendix
    \renewcommand\thesection{S}
    \beginsupplement
\section{Supplementary information}
    \subsection{Second example chip design}\label{SI_Chipdesign}
    \begin{figure}[htb]
    \centering
    \includegraphics[width=\linewidth]{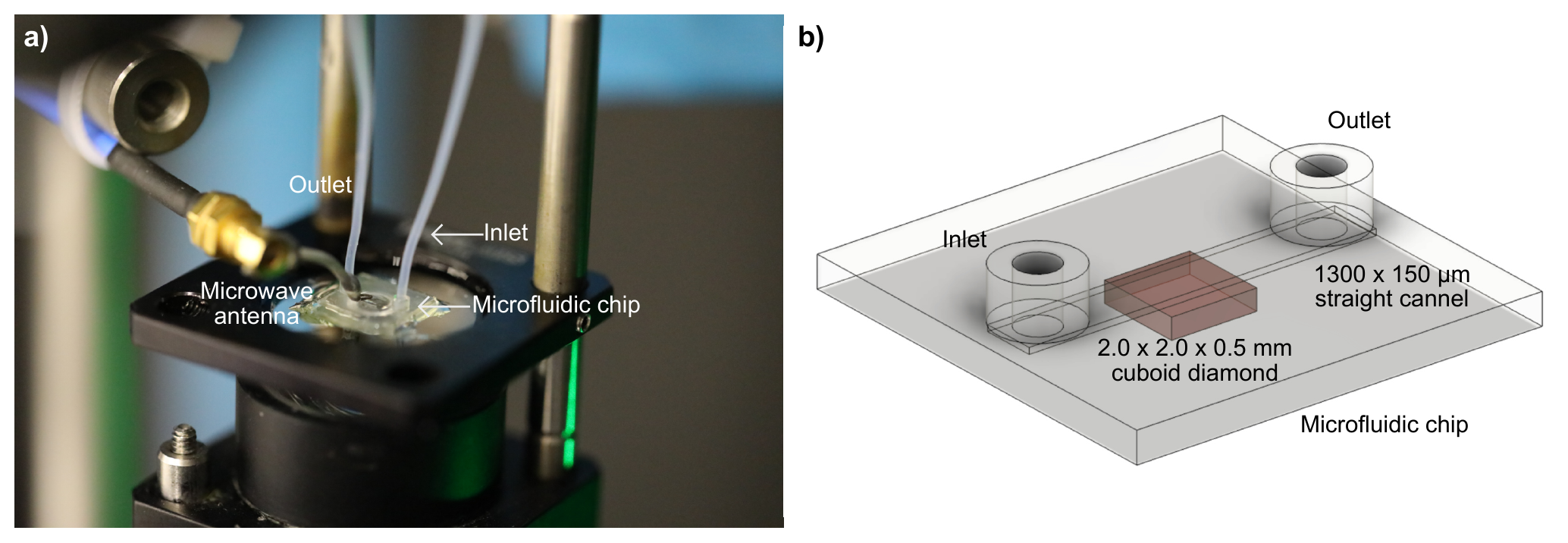}
    \caption{\textbf{Second microfluidic chip design.} \textbf{a)} Photograph of the second microfuidic chip design inside a NV center experiment. \textbf{b)} CAD image of the second chip design which incorporates a 2.0 x 2.0 x 0.5~mm cubiod-shaped diamond and a 1,300 x 150~\si{\micro\meter} straight example channel. The inlet and outlet ports in this design are on top of the channel to allow easy optical access through the diamond by, \textit{e.g.}, solid-immersion lenses (SILs) or microscopy objectives.}
    \label{Figure_SI1}
    \end{figure}
The second microfluidic chip design fits (see Fig.~\ref{Figure_SI1}) standard 2.0 x 2.0 x 0.5~mm NV center-doped cuboid diamonds, demonstrating that the microfluidic quantum sensing platform can easily be adapted to different diamond shapes and sizes. Furthermore, in this design, the inlet and outlet ports for the microfluidics are placed on top of the chip to facilitate optical access for NV center excitation and read-out from the bottom through the diamond. This geometry has the drawback that a total internal reflection geometry is more difficult to realize but allows for a more rigid chip design where no cut-out for the excitation pathway is needed (compare main text Fig.~\ref{Figure_3}\textcolor{blue}{a}). We envision applications using confocal microscopy or other nanoscale experiments, where photodegradation of biological or chemical samples is less of a concern. In this example, the microfluidic channel is 2,000 x 1,300 x 150~\si{\micro\meter}, which can be fully adapted to more complicated structures. A technical drawing of the microfluidic chip is depicted in Fig.~\ref{Figure_SI2}\textcolor{blue}{c}).
    \subsection{Technical drawings}\label{SI_TechDrawings}
\begin{figure}[htb]
        \centering
        \includegraphics[width=\linewidth]{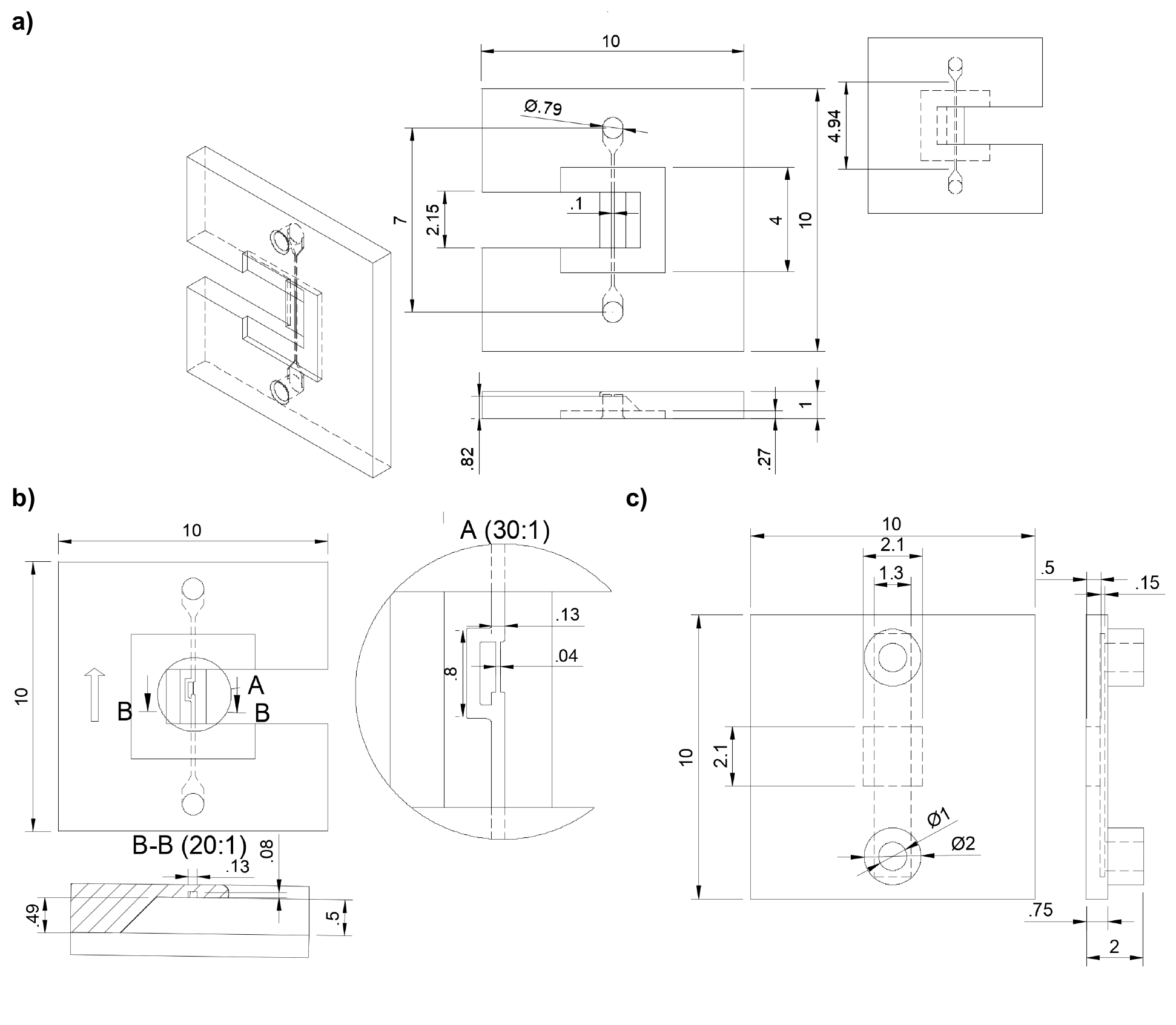}
        \caption{\textbf{Technical drawings of example microfluidic chips.} 
        \textbf{a)} Simplified technical drawing of the base microfluidic chip with a 100 x 80~\si{\micro\meter} straight channel (see main text Fig.~\ref{Figure_3}\textcolor{blue}{a}), \textbf{b)} of a modified microfluidic chip with a fluidic bypass channel (see main text Fig.~\ref{Figure_3}\textcolor{blue}{b}), and
        \textbf{c)} of the second microfluidic chip design shown in SI Section~\ref{Figure_SI1} (see Fig.~\ref{Figure_SI1}).
        }
        \label{Figure_SI2}
    \end{figure}

Fig.~\ref{Figure_SI2} depicts simplified technical drawings of several example microfluidic chips that have been used for the microfluidic quantum sensing platform. The microfluidic chip designs are attached in the Supplementary Data as .STL files.
    \subsection{Optimization of the NV layer thickness for NV-NMR within our microfluidic chips}\label{SI_Sec_layeroptimization}
    \begin{figure}[htb]
        \centering
        \includegraphics[width=\linewidth]{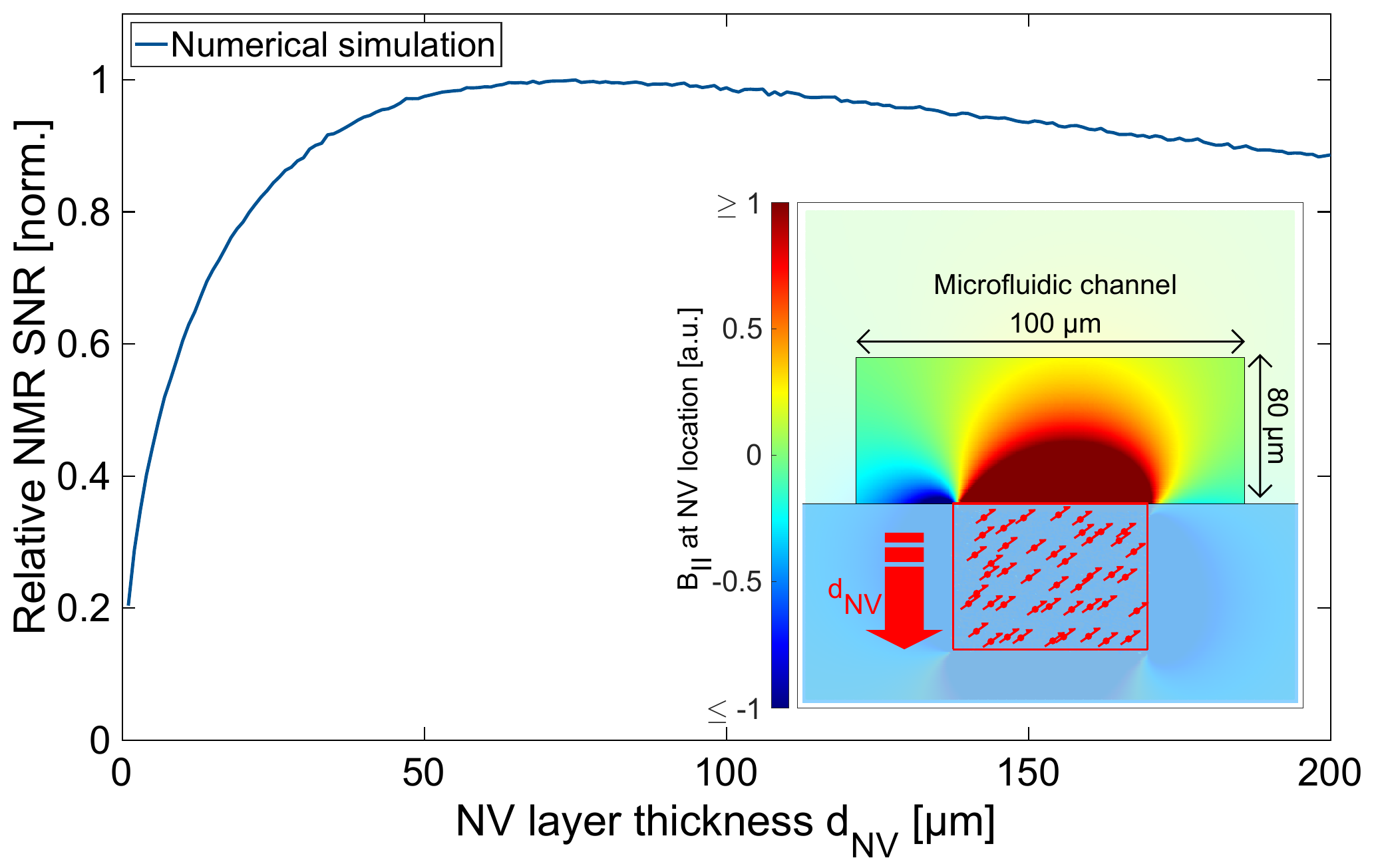}
        \caption{\textbf{Optimization of the NV layer thickness to the microfluidic dimensions for microscale NV-NMR.}
        \textbf{Inset:} Visualization of the NMR signal origin for a microscale NV-NMR experiment with a cylindrical NV center (red arrows) volume, \textit{i.e.}, sensor volume, of 45~\si{\micro\meter} diameter and variable height (\si{d_{NV}}) at the surface of the diamond. The sample spins are contained in a 1,000 x 100 x 80~\si{\micro\meter} microfluidic channel. For each sample spin position, the dipolar field within the detection volume is projected along the magnetic bias field B$_0$ and averaged. Red areas depict sample spin positions with overall positive signal contribution, and the blue regions sample spin positions with overall negative signal contribution, respectively. \textbf{Graph:} Normalized signal \si{B_{II}\sqrt{d_{NV}}}, \textit{i.e.} the relative sensitivity of the NV-NMR experiment, against the thickness of NV ensemble layer \si{d_{NV}} according to a numerical simulation using the Monte Carlo method.
        }
        \label{Figure_SI3}
    \end{figure}
 The microscale NV-NMR signal has a strong volume and geometry dependence~\cite{Glenn2018-hp}. In order to choose the NV-layer thickness for the highest sensitivity, we follow the approach according to Bruckmaier \textit{et al.}~\cite{Bruckmaier2021}. Using Monte Carlo simulation, we estimate the NMR signal strength at the NV center position (red arrows, see Fig.~\ref{Figure_SI3}), caused by the dipolar magnetic fields of nuclear sample spins (protons). These spins are confined in our 1,000 x 100 x 80~\si{\micro\meter} microfluidic channel (see Fig.~\ref{Figure_3}\textcolor{blue}{a}). The cylindrical NV center volume, \textit{i.e.}, the sensor volume, is 45~\si{\micro\meter} in diameter (typical value in microscale NV-NMR experiments, see Fig.~\ref{Figure_3}\textcolor{blue}{d}) and has a variable height (\si{d_{NV}}) form the surface of the diamond. For each sample spin position, the dipolar field within the detection volume is projected along the NV orientation, (\textit{i.e.}, the magnetic bias field B$_0$ direction). Each point was averaged 10,000 times for this simulation with 40 randomly selected NV center positions within the volume and 32,000 randomly selected sample spins within the microfluidic channel for each average. While the average signal size of each of our NV center within the ensemble decreases with the depth \si{d_{NV}}, the sensitivity (relative NMR SNR) reaches a maximum at $\sim$~50 - 70~\si{\micro\meter} due to the increasing number of NV centers contributing to the measurement (see Fig.~\ref{Figure_SI3}). For that reason, we grew a 50~\si{\micro\meter} thick NV-layer for our microscale NV-NMR experiments within the channel, yielding the highest sensitivity for our microfluidic chip dimensions.  
\end{document}